\documentclass[aip,pof,preprint]{revtex4-1}

\usepackage{epsfig}
\usepackage{amsmath}
\usepackage{amssymb}
\usepackage{graphicx}
\usepackage{dcolumn}
\usepackage{bm}
\usepackage{url} 

\def\({\left(}
\def\){\right)}
\def\xp{x^{\prime}}
\def\zp{z^{\prime}}
\def\dxp{\!\mathrm{d}x^{\prime}\, }
\def\dzp{\!\mathrm{d}z^{\prime}\,}
\def\dt{\!\mathrm{d}t\,}

\begin{document}


\title{Experimental determination of radiated internal wave power without pressure field data}

\author{Frank M. Lee}
\affiliation{Physics Department and Institute for Fusion Studies, 
The University of Texas at Austin, Austin, TX 78712--1192}

\author{M. S. Paoletti\footnote{Present Address:  Applied Physics Laboratory, Johns Hopkins University, Baltimore, MD 21218}}
\affiliation{Physics Department, The University of Texas at Austin, Austin, TX 78712--1192}

\author{Harry L. Swinney}
\affiliation{Physics Department, The University of Texas at Austin, Austin, TX 78712--1192}

\author{P. J. Morrison}
\affiliation{Physics Department and Institute for Fusion Studies, The University of Texas at Austin, 
 Austin, TX 78712--1192}

\date{\today}

\begin{abstract}
We present a method to determine, using only velocity field data, the time-averaged energy flux $\left<\bm{J}\right>$ and total radiated power $P$ for two-dimensional internal gravity waves.   Both $\left<\bm{J}\right>$ and $P$ are determined from expressions involving only a scalar function, the stream function $\psi$. We test the method using data from a direct numerical simulation for tidal flow of a stratified fluid past a knife edge.   The results for the radiated internal wave power given by the stream function method agree to within 0.5\% with results obtained using pressure and velocity data from the numerical simulation.  The results for the radiated power computed from the stream function agree well with power computed from the velocity and pressure if the starting point for the stream function computation is on a solid boundary, but if a boundary point is not available, care must be taken to choose an appropriate starting point.  We also test the stream function method by applying it to laboratory data for tidal flow past a knife edge, and the results are found to agree with the direct numerical simulation.  Supplementary Material includes a Matlab code with a graphical user interface (GUI) that can be used to compute the energy flux and power from any two-dimensional velocity field data. 
\end{abstract}

\pacs{Valid PACS appear here}
\maketitle

\section{Introduction} \label{sec:Introduction}

Internal waves transport momentum and energy in stably stratified fluids as propagating disturbances that are restored by buoyancy forces.  The thermohaline circulation in the ocean stems at least in part from the conversion of energy in large scale tidal and rotational motions into internal waves that eventually break and deposit their energy into gravitational potential energy through irreversible, small-scale mixing.\cite{munk98,wunsch04}  To determine the role that internal waves play in global ocean mixing, it is important to understand the power present in the internal wave field.  The average power radiated by an internal wave beam through a closed surface $S$ is given by,
\begin{equation}
\label{eq:P}
{P} = \int_S \!d^2x\,   \langle\boldsymbol{J}\rangle\cdot \hat{\boldsymbol{n}}  = \int_S\!d^2x\,  \left< p \boldsymbol{v} \right> \cdot \hat{\boldsymbol{n}}\,, 
\end{equation}
where $\boldsymbol{J} =  p\boldsymbol{v}$ is the baroclinic energy flux, $p$ is the perturbed pressure field, $\boldsymbol{v}$ denotes the velocity perturbation, brackets $\left<\  \right>$ indicate a temporal average over an integer number of tidal periods, and $\hat{\boldsymbol{n}}$ is a unit vector normal to the surface $S$.  

Theoretical\cite{robinson69,baines73,baines74,bell75,balmforth02,smith02,smith03,khatiwala03,stlaurent03,lorenzo06,nycander06,petrelis06,garrett07,griffiths07,balmforth09,echeverri10,zarroug10} and numerical studies\cite{holloway99,lamb04,legg04a,legg04b,niwa04,munroe05,zilberman09,king09,king10,qian10,gayen10,gayen11a,gayen11b,rapaka13,dettner13} have sought to determine the efficiency of the conversion of energy from tidal and rotational motions over bottom topography into radiated internal waves, but laboratory and field measurements of internal wave power remain scarce,  owing to the difficulty in simultaneously measuring the perturbed pressure and velocity fields.  Particle image velocimetry\cite{adrian91} has been used in laboratory studies of internal waves to characterize the velocity fields,\cite{guo03,zhang07,zhang08,echeverri09,king09,king10,paoletti12,paoletti13} and synthetic schlieren has been used in a few studies to measure density perturbations averaged along the line of sight;\cite{aguilar06,gostiaux07,clark10} however, measurements of the accompanying pressure fields have not been made owing to technical challenges in doing so.  To circumvent the difficulty in measuring the pressure field to obtain the energy flux of two-dimensional internal waves, \citet{echeverri09} decomposed their experimental velocity fields into the three lowest vertical modes,\cite{sutherland} which could be used to estimate the internal wave energy flux using Eq.\ (2.31) from \citet{petrelis06}.  The advantage of this method is that it removes the need for determining the pressure field.  However, the vertical modal  analysis assumes that the result of linear, inviscid theory\cite{petrelis06} is applicable, requires knowledge of the velocity field along the entire fluid depth, and becomes less reliable for internal wave beams characterized by higher modes.  In other laboratory experiments, \citet{aguilar06} applied results from linear theory to convert two-dimensional measurements of vertical displacements, measured by the synthetic schlieren technique, into estimates of the momentum flux.  Building upon this work, \citet{clark10} used the Boussinesq polarization relations and the spatial structure of internal wave fields (given by Fourier modes of their measurements of vertical displacements) to estimate the internal wave power radiated by internal wave beams.  In those studies, measurements of modal amplitudes were coupled with linear theory to estimate the energy flux.

The need to determine the pressure field for a given fluid flow is not unique to internal wave studies.  Recent technical developments have allowed for measurements of the velocity field to be coupled with the governing equations to determine the dynamic pressure field, as reviewed by \citet{oudheusden13}.  \citet{jakobsen97} used a four-CCD-camera system to determine the acceleration and pressure for surface waves.  \citet{jensen01} used two separate cameras to measure the velocity field of the same region of the flow at two closely spaced instants in time to determine the acceleration field of water waves, which can be used to obtain the pressure field.    By using high temporal resolution particle image velocimetry, \citet{baur99} measured the local pressure reduction in vortices produced by shear layers behind an obstacle.  \citet{liu06} used a four-exposure particle image velocimetry system to measure the pressure distribution in a turbulent cavity flow by spatially integrating their measurements of the material acceleration.  These techniques all require either multiple cameras or high temporal resolution to allow for measurements of the material acceleration, which together with the Navier-Stokes equation can be used to solve for the pressure field.  

\begin{figure}
\includegraphics[width=\textwidth]{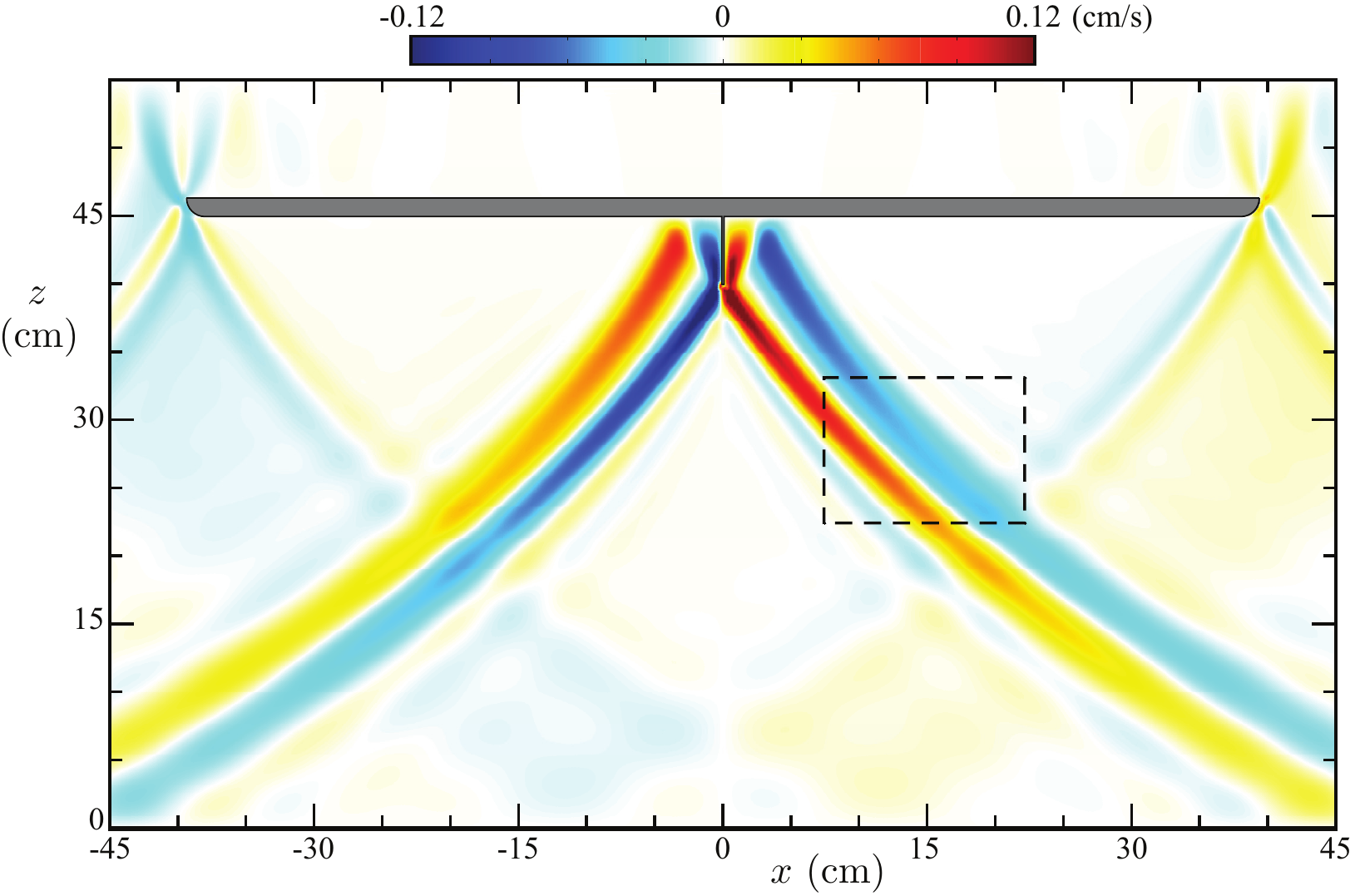}
\caption{This snapshot of the vertical component of the velocity field (color), computed in numerical simulations for the same conditions as our laboratory experiments, reveals internal wave beams generated by knife edge topography (located at the top) that oscillates about $x = 0$; weaker internal waves are  generated by the ends of the gray base plate.  This numerical simulation mimics the finite-size effects present in the experiments, where waves reflect from the top and bottom boundaries but are damped for $|x| > 45$~cm. The dashed box shows the location of the experimental measurements of the velocity field. The internal wave beams bend because the buoyancy frequency $N(z)$ varies exponentially with $z$, as described in Sec.~\ref{subsec:expt_techniques}.   This snapshot is at  time $t/T=7.525$ after initiation of the oscillations, where $T = 2\pi/\omega = 6.98$~s is the oscillation period for a tidal excursion with amplitude $A = 1$~mm.}
\label{fig:wsim_lab}
\end{figure}

Here, we present a method for determining the energy flux and radiated power for propagating internal waves without any knowledge of the pressure field, and we apply the method to results from direct numerical simulations and from laboratory data for tidal flow past a knife edge, for the geometry shown in Fig.\ \ref{fig:wsim_lab}.  To circumvent the need for the perturbed pressure field in Eq.\ \eqref{eq:P}, we assume that the velocity field is predominantly two-dimensional, as has been the case in many internal wave studies.  A two-dimensional velocity field with zero divergence can be expressed in terms of a scalar field, the stream function $\psi$, where
\begin{equation}
\label{eq:psi_v_def}
\boldsymbol{v} = u\hat{\boldsymbol{x}} + w\hat{\boldsymbol{z}} =  -\frac{\partial \psi}{\partial z} \hat{\boldsymbol{x}} + \frac{\partial \psi}{\partial x} \hat{\boldsymbol{z}}.
\end{equation}
It is straightforward to derive  an  expression for the energy flux for two-dimensional internal waves in terms of the stream function (see,  e.g., Refs.~\onlinecite{balmforth02,smith03}).  We use such an expression to compute the radiated internal wave power from particle imaging velocimetry measurements for tidal flow over a knife edge ridge.  This wave power is compared with that obtained from companion numerical simulations of the Navier-Stokes equations, where the power can be directly computed from Eq.\ \eqref{eq:P}.  

The theory behind our approach is presented in Sec.~\ref{sec:Theory} and our methods are described in Sec.~\ref{sec:Methods}. In Sec.~\ref{sec:Results} we show, using data from a numerical simulation of the Navier-Stokes equations, that the internal wave power obtained using the stream function method agrees with that obtained from pressure and velocity field data, provided that appropriate attention is given to the choice of the starting point for the stream function calculation. We then apply the stream function method to calculate internal wave power for laboratory data. The paper ends with a discussion in Sec.~\ref{sec:Discussion} and our conclusions in Sec.~\ref{sec:Conclusions}. An appendix is included as a guide to the supplementary material. 


\section{Theory} \label{sec:Theory}
The derivation of the equations that describe internal waves starts with the compressible Navier-Stokes equations,\begin{eqnarray}
&&
\rho \left[ \frac{\partial \bm{v}}{\partial t} + \bm{v} \cdot \nabla \bm{v} \right] = -\nabla p + \mu \nabla^2 \bm{v} +\rho \,\bm{a}
\\
&&
\frac{1}{\rho} \left[ \frac{\partial \rho}{ \partial t} +  \bm{v} \cdot \nabla \rho \right] + \nabla \cdot \bm{v} = 0.
\end{eqnarray}
where $\rho$ is the density, $\bm{v}$ is the velocity, $\bm{a}$ is the acceleration due to external forces, and $\mu$ is the coefficient of viscosity.  Then, assuming that a linear approximation about a background equilibrium state with a density  only  dependent on the height is appropriate, that the fluid is inviscid and incompressible, and that the only external force is that due to gravity,  we obtain  the following set of linearized 2-dimensional equations for internal waves: 
\begin{eqnarray}
&&\frac{\partial u}{\partial t} = -\frac{1}{\rho_0} \frac{\partial p}{\partial x}\,,  
\qquad
\frac{\partial w}{\partial t} = -\frac{1}{\rho_0} \frac{\partial p}{\partial z} - \frac{\rho}{\rho_0} g\,, \label{dudt}    \\
&&
\frac{\partial \rho}{\partial t} = \frac{N^2 \, \rho_0}{g} w\,,  \qquad
\frac{\partial u}{\partial x} + \frac{\partial w}{\partial z} = 0\,, 
 \label{div1}
\end{eqnarray}
where $x$ and $z$ are the horizontal and vertical  coordinates, respectively, $u$ and $w$ are the corresponding components of the velocity,  and $p$ and $\rho$ are the pressure and density perturbations away from a hydrostatic background described by  $\rho_0 = \rho_0(z)$,  with $g$ the acceleration due to gravity and $N$  the buoyancy frequency,
\begin{align}
N = \sqrt{ \frac{-g}{\rho_0} \frac{\partial \rho_0}{\partial z} }. 
\label{N}
\end{align}
When the density variations are weak enough so as to not significantly affect inertial terms, it is common  to replace $\rho_0(z)$ by a constant value denoted $\rho_{00}$, while $N$ retains a $z$-dependence.  This procedure is referred to as  the Boussinesq approximation.  The flux and  power formulas we derive will be valid both with and without this approximation. 

In the Boussinesq approximation with constant $N$,  one can assume  traveling wave solutions, e.g., $\rho \sim  \, e^{i(k_x x + k_z z - \omega t)}$, and similarly for $p$, $u$, and $w$, leading  to  the usual internal wave dispersion relation, $\omega^2 = N^2 ( 1- k_z^2/|k|^2 )$ or $\omega  = N \sin(\theta)$. Here, $\theta$ is the angle of $\bm{k}$ with respect to the $z$-axis, the vertical. 

For  2-dimensional incompressible flow, the perturbation velocity components can be expressed in terms of a stream function $\psi$, as in (\ref{eq:psi_v_def}).  Then, using \eqref{eq:psi_v_def} and neglecting viscous dissipation,  the equations of motion \eqref{dudt} and \eqref{div1} imply energy conservation   as follows:
\begin{align}
\nabla \cdot \bm{J} = -\frac{\partial E}{\partial t} :=& 
-\frac{\partial}{\partial t} \left[ \frac{\rho_0}{2} ( u^2 + w^2 ) - \frac{\rho^2 g}{2 \partial \rho_0/ \partial z} \right]\nonumber \\
                                                     =& \, u \, \frac{\partial p}{\partial x} +  w \, \frac{\partial p}{\partial z}
                   = -\frac{\partial \psi}{\partial z} \, \frac{\partial p}{\partial x} 
                   +  \frac{\partial \psi}{\partial x} \, \frac{\partial p}{\partial z}\,. 
\label{divj}                                  
\end{align}
where  $\bm{J}$ is the energy flux.    Equation  \eqref{divj} implies various  solutions for $\bm{J}$, e.g., 
\begin{equation}
\bm{J}_p =  -\frac{\partial \psi}{\partial z} \, p \,\bm{\hat{x}} + \frac{\partial \psi}{\partial x} \, p \,\bm{\hat{z}}  
= p(u  \,\bm{\hat{x}} + w  \,\bm{\hat{z}}),   \label{j1}
\end{equation}
or
\begin{equation}
\bm{J}_{\psi} = \psi \,\left( \frac{\partial p}{\partial z} \,\bm{\hat{x}} - 
 \, \frac{\partial p}{\partial x} \,\bm{\hat{z}} \right), 
 \label{j2}
\end{equation}
where $\bm{J}_p$ and $\bm{J}_{\psi}$ differ by a gauge condition, 
\begin{equation}
\bm{J}_p = \bm{J}_{\psi} + {\nabla} \times (\psi p \, \bm{\hat{y}}). 
\end{equation}
The form of  Eq.~\eqref{j1},  $\bm{J}_p= p\bm{v}$, is the commonly used expression for the energy flux.  However, we will present a form obtained from (\ref{j2})
with  further manipulation. While the  form  of \eqref{j1}  requires both the velocity and pressure fields over time, the  form we use will depend ultimately only on the velocity field.

We assume the stream function can be written as 
\begin{align}
\psi(x,z,t) = Re\{ e^{-i\omega t} \varphi(x,z) \}\,,  \label{psiseparation}
\end{align}
where 
$\omega$ is the angular frequency of the internal waves and $\varphi$ is the spatially dependent amplitude that is in general complex. Using \eqref{dudt}, \eqref{div1}, \eqref{j2}, and \eqref{psiseparation},  the following expression for the time-averaged energy flux is obtained:
\begin{align}
 \langle{\bm{J}}_{\psi}\rangle	:=& \frac{i  \rho_0}{4 \omega} \left[ (N^2 - \omega^2)\left(\varphi \, \frac{\partial \varphi^{*}}{\partial x} - \varphi^{*} \, \frac{\partial \varphi}{\partial x} \right)\,\bm{\hat{x}} -\omega^2\left( \varphi \, \frac{\partial \varphi^{*}}{\partial z} - \varphi^{*} \, \frac{\partial \varphi}{\partial z} \right)\,\bm{\hat{z}} \right]. \label{j2tavg}
\end{align}
To obtain the result in the Boussinesq approximation one simply replaces $\rho_0$ in this expression by the constant $\rho_{00}$.

The functions $\varphi$ and $\varphi^{*}$ can be found from the stream function $\psi$,  which in turn can be obtained from the velocity field.  Thus,  the  energy flux expression  $\langle{\bm{J}}_{\psi}\rangle$  does not require any knowledge of the pressure perturbations,  in contrast to the standard form of \eqref{j1}, which when averaged over a period becomes 
\begin{align}
\langle{\bm{J}}_{p}\rangle :=& \frac1{T} \int_{t_0}^{t_0 + T} \hspace{-.3cm} dt \,\,   p 
\bm{v} 
 = 
\frac{1}{4} \big[
 (\mathring{u} \mathring{p}^{*} + \mathring{u}^{*} \mathring{p} ) \, \bm{\hat{x}} +  (\mathring{w} \mathring{p}^{*} + \mathring{w}^{*} \mathring{p} ) \,\bm{\hat{z}}
 \big], \label{j1tavg}
\end{align}
with $T=2 \pi/\omega$,  $u(x,z,t) = Re\{ e^{-i \omega t} \, \mathring{u}(x,z) \}$,  and similar expressions for  $w$ and $p$ (and the complex conjugates $u^*, w^*$, and $p^*$) written in terms of their amplitudes.  
 
Our calculation of $\langle{\bm{J}}_{\psi}\rangle$ for the time-averaged energy flux is essentially the same as that in Refs.~\onlinecite{balmforth02,smith03}, although these authors show an explicit dependence on the tidal velocity amplitude. They also use the Boussinesq approximation  and, in addition, Ref.~\onlinecite{smith03} makes an hydrostatic approximation;  more significantly,   those authors did not use  expression (\ref{j2tavg})  to interpret experimental data in the manner we describe below.      Note,  since  the two energy fluxes of Eqs.~\eqref{j1} and \eqref{j2}  differ  by a curl (the gauge term),  the total power given by
\begin{equation}
P = \int_{\partial V} \! d^2x \,  \langle\bm{J}\rangle \cdot \hat{\bm{n}} 
= \int_{V} \!d^3x\, \nabla \cdot \langle\bm{J}\rangle\,, 
\label{eq:J_integrals}
\end{equation}
where ${\partial V}$ is the surface bounding a volume $V$, will be identical when either   $\bm{J}_{\psi}$ or $\bm{J}_{p}$ is inserted.

Thus,  only the perturbation velocity field is needed to compute the power produced by topography in the form of internal waves.  The caveat is that because the  internal wave equations were used to derive the time-averaged flux fields, the result would only be correct for a system that is dominated by internal waves. Additionally, the equations of motion used were linearized and inviscid, which means if there is a significant presence of higher-order harmonics or appreciable amounts of damping, the results might not be reliable.    However, our simulations indicate the method is  robust to the inclusion of dissipation.   Also, because of the temporal periodicity assumption of \eqref{psiseparation}, the system should ideally  be in a steady state or close to it. Thus,  even though we do not require knowledge of either the perturbation pressure or perturbation density field, use of $\langle{\bm{J}}_{\psi}\rangle$ narrows the scope of applicability to linear internal waves near  a steady state with small  damping.  However, because the method does not require data from these two perturbation fields, obtaining the time-averaged energy flux of internal waves in the ocean is possible. Also,  the details of the topography itself do not matter, as long as the velocity fields are solutions of the internal wave equations.

\section{Methods}\label{sec:Methods}

This section describes our methods:  the computational algorithm for the flux in Sec.~\ref{subsec:Computational_Flux}, the numerical simulations of the Navier-Stokes equations in Sec.~\ref{subsec:simulations}, and the experimental geometry and techniques in Sec.~\ref{subsec:expt_techniques}, which also shows that the simulation and experimental results are in good agreement. 

\subsection{Computational Algorithm for the Flux} 
\label{subsec:Computational_Flux}

In this section we use data from simulations with a constant  buoyancy frequency (domain 2 (Grid II) of  Sec.~II.B) in order to validate our method by comparison with analytic theory.  A snapshot of the results of the simulations for tidal flow past the knife edge topography (discussed in the next subsection) is  illustrated in Fig.~\ref{fig:wsim_lab}.  In  order to compute the energy flux from only the velocity field using \eqref{j2tavg}, we must first obtain the stream function $\psi$ by inverting the relations of Eq.~\eqref{eq:psi_v_def}.  This can be done by using the incompressibility condition \eqref{div1} and integrating  \eqref{eq:psi_v_def}, resulting in
\begin{equation}
\psi\(x,z,t\) = \int_{x_0}^{x} \!\mathrm{d}x^{\prime}\, w\(x^{\prime},z_0,t\) 
- \int_{z_0}^{z}\! \mathrm{d}z^{\prime} \,   u\(x,z^{\prime},t\) + \psi\(x_0,z_0,t\), 
\label{eq:psi1}
\end{equation}
where $\(x_0,z_0\)$ is the starting point for the integration, and $\psi\(x_0,z_0,t\)$ is an arbitrary integration constant, which we set to zero for our calculations.  We discuss the importance of properly choosing the  point  $\(x_0,z_0\)$ in Sec.~\ref{subsec:starting_pt}.  The integral from $\(x_0,z_0\)$ to $\(x,z\)$ is given in Eq.~\eqref{eq:psi1} by first integrating  the vertical velocity field along the $x$-direction and then  integrating the horizontal velocity field in the $z$-direction.

\begin{figure}[htb]
\includegraphics[width=\textwidth]{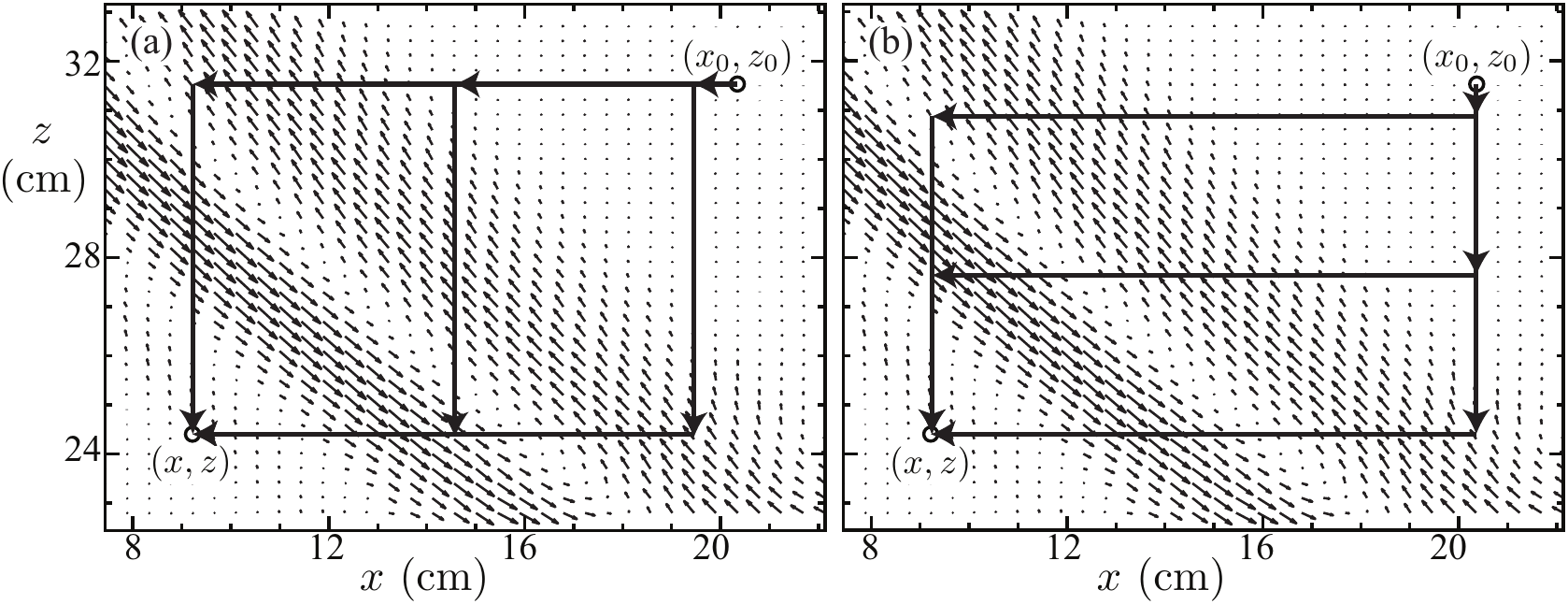}
\caption{The stream function at a point $(x,z)$ is determined by averaging the velocities integrated along paths that (a) first travel horizontally from the starting point $(x_0, z_0)$ towards the point $(x,z)$, then vertically, and then horizontally again, as well as (b) paths that first travel vertically, then horizontally, and then vertically again.  The velocity component perpendicular to the path appears in each integrand, and all of the experimental or  computational grid points in the box with corners at $(x_0, z_0)$ and $(x,z)$ are used.  The conditions for these data are given in the caption for Fig.\ \ref{fig:wsim_lab}. }
\label{fig:paths} 
\end{figure}

Since the stream function serves as a scalar potential for a conjugate velocity field,  its  values are theoretically  independent of the path of integration.  Therefore, we can also compute the stream function in the following manner:
\begin{equation}
\psi\(x,z,t\) = -\int_{z_0}^{z} \!\mathrm{d}z^{\prime} \, u\(x_0,z^{\prime},t\) + \int_{x_0}^{x} \! \mathrm{d}x^{\prime} \, w\(x^{\prime},z,t\) +  \psi\(x_0,z_0,t\).
\label{eq:psi2}
\end{equation}
In this case, the stream function is obtained by first integrating the horizontal velocities along a vertical path, and then integrating the vertical velocities along a horizontal path.  Indeed, we are not restricted to these two specific paths as any path between the points $\(x_0,z_0\)$ and $\(x,z\)$ can be used to compute the stream function.  Thus, we can use any collection of paths that first travel along the grid horizontally, then vertically, and finally horizontally again, as shown in Fig.\ \ref{fig:paths}(a).  Such paths of integration are given by
\begin{align}
\psi\(x,z,t\) = & \int_{x_0}^{x_i} \dxp  w\(\xp,z_0,t\) 
- \int_{z_0}^{z}  \dzp u\(x_i,\zp,t\) \nonumber\\
   &+ \int_{x_i}^{x} \dxp  w\(\xp,z,t\)+ \psi\(x_0,z_0,t\), 
\label{eq:psi3}
\end{align}
where $x_i$ is any point between $x_0$ and $x$. We can also take paths that first travel vertically, then horizontally, and then vertically again, as shown by Fig.~\ref{fig:paths}(b) (see Fig.~\ref{fig:paths}(b)),
\begin{align}
\psi\(x,z,t\) =& -\int_{z_0}^{z_i}\dzp   u\(x_0,\zp,t\)+ \int_{x_0}^{x} \dxp  w\(\xp,z_i,t\)
\nonumber \\
                              & -\int_{z_i}^{z}  \dzp u\(x,\zp,t\) + \psi\(x_0,z_0,t\), 
\label{eq:psi4}
\end{align}
where $z_i$ is any point between $z_0$ and $z$.

\begin{figure}[htb]
\includegraphics[width=\textwidth]{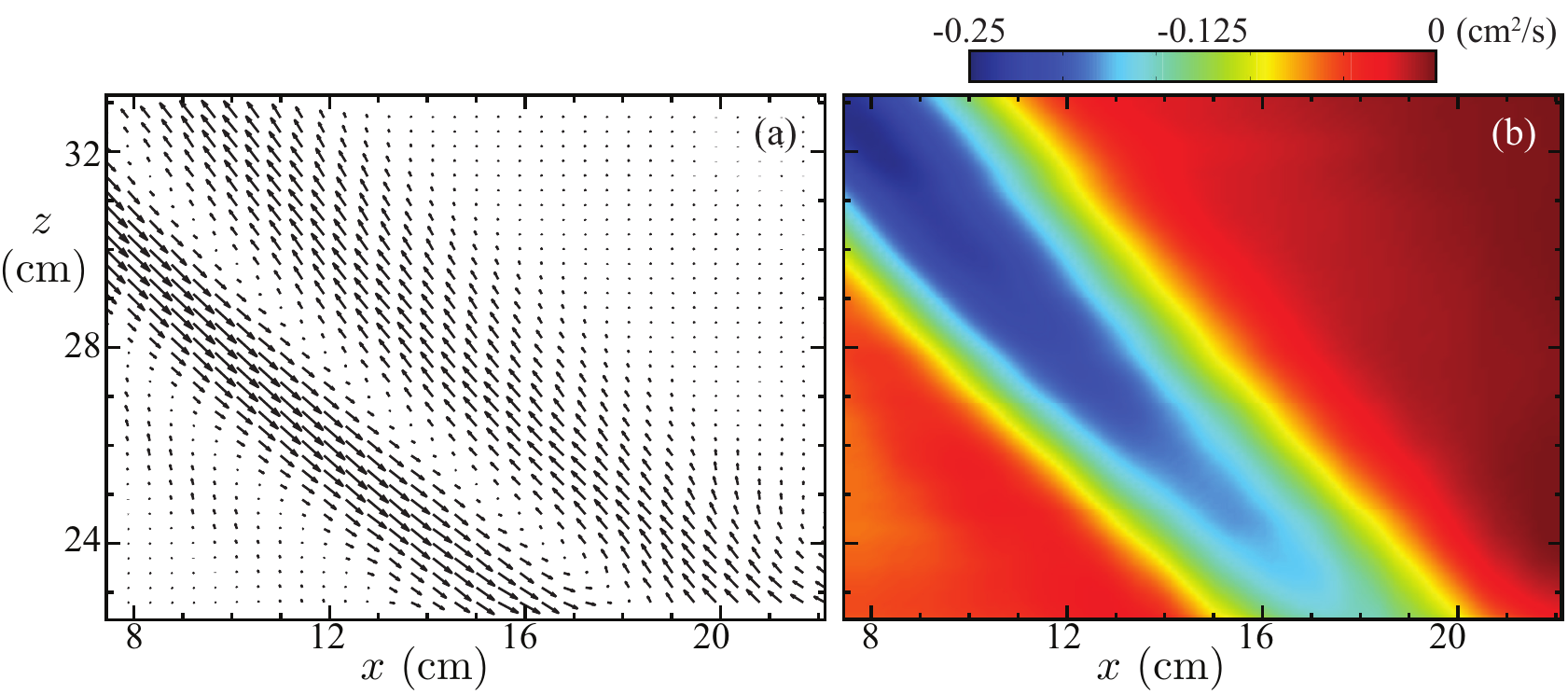}
\caption{(a) A snapshot of the 2-dimensional velocity field used to compute (using Eqs.~(\ref{eq:psi3}) and (\ref{eq:psi4})) (b) the  corresponding stream function $\psi(x,z,t)$ with the top right corner as the starting point $\(x_0,z_0\)$. The conditions for these data are given in the caption for Fig.\ \ref{fig:wsim_lab}.}
\label{fig:vfield_and_stream} 
\end{figure}

Statistical errors in the  stream function can be minimized by computing the average value for all possible paths for the grid used in the simulation or experiment of the types shown in Fig.\ \ref{fig:paths} (as given by Eqs.\ (\ref{eq:psi3}) and (\ref{eq:psi4})) between the starting point $\(x_0,z_0\)$ and the point of interest $\(x,z\)$.  Figure \ref{fig:vfield_and_stream} shows a snapshot of our experimental velocity data and the corresponding stream function.  We find that stream function values computed from our experimental data using only the two paths defined by Eqs.~(\ref{eq:psi1}) and (\ref{eq:psi2}) differ by less than 1\% from more computationally intensive multi-path method indicated in Fig.\ \ref{fig:paths}. 
However, the computationally more expensive multi-path method would be preferable for noisy data. Optimization of  the multi-path method could be pursued, but we do not do this here.

In principle the choice of the starting point $\(x_0,z_0\)$ should not affect the computed internal wave power.  However, in practice $\(x_0,z_0\)$ must be chosen carefully, because of the arbitrary integration constant $\psi\(x_0,z_0,t\)$.  \citet{balmforth02} effectively chose a starting point along the boundary, where they could specify $\psi\(x_0,z_0,t\) = \mathrm{constant}$ owing to the no-penetration boundary condition.  We show in Sec.~\ref{subsec:starting_pt} that choosing a starting point along or near a boundary is the best choice.  If the experimental velocity field does not contain points near a boundary, then we find that a starting point away from the internal wave beams also works well. For example, a point in the upper right corner of Fig.\ \ref{fig:vfield_and_stream} is satisfactory.  

After determining the stream function for a tidal period $T$, the real and imaginary parts of the  field $\varphi(x,z)$ must be determined by inverting Eq.\ \eqref{psiseparation}.  Specifically, we have
\begin{equation}
 \varphi(x,z) = \frac{2}{ T} \int_{t_0}^{t_0 + T}\! \dt \, {\psi(x,z,t)\,  e^{i \omega t}}\,.
 \label{rephi} \\
\end{equation} 
The derivatives of $\varphi$ that appear in Eq.\ \eqref{j2tavg} are determined by moving them  into the integrand and using the relations \eqref{eq:psi_v_def}:
\begin{align}
 \frac{\partial \varphi}{\partial x}  =& \frac{2}{T} \int_{t_0}^{t_0 + T}\!\dt \,
 {w(x,z,t) \, e^{i \omega t}}\,,  
 \label{rephix} \\
 \frac{\partial \varphi}{\partial z}   =& - \frac{2}{T} \int_{t_0}^{t_0 + T} \!\dt\,  {u(x,z,t)\,  e^{i \omega t}}\,.
\label{imphiz}
\end{align} 
The field $\varphi$ and its derivatives can then be used in conjunction with  known 
background density profile $\rho_{0}(z)$, buoyancy frequency profile $N(z)$, and tidal frequency $\omega$ to determine the tidally-averaged energy flux field by Eq.\ \eqref{j2tavg} and the radiated power from Eq.\ \eqref{eq:P}.

\subsection{Navier-Stokes numerical simulations} 
\label{subsec:simulations}

We numerically simulate the generation and propagation of internal waves by tidal flow of a stratified fluid over a knife edge ridge by solving the Navier-Stokes equations in the Boussinesq approximation using the code CDP-2.4.\citep{ham04}  This code is a parallel, unstructured, finite-volume-based solver modeled after the algorithm of \citet{mahesh04}; all subgrid scale modeling is disabled.  By using a fractional-step time-marching scheme and multiple implicit schemes for the spatial operators, \citep{ham06} the code achieves second-order accuracy in both space and time. The following equations are solved for the density $\rho$, pressure $p$, and velocity field $\boldsymbol{v} = (u(x,z), w(x,z))$:
\begin{eqnarray}
\label{eq:NS}
&&\frac{\partial{\boldsymbol{v}}}{\partial{t}} + \boldsymbol{v} \boldsymbol{\cdot} {\nabla}\boldsymbol{v} = -\frac{1}{\rho_{00}}\nabla p - \frac{g\rho}{\rho_{00}}\hat{\boldsymbol{z}} + \nu \nabla^2 \boldsymbol{v} + \frac{F_{\mathrm{tide}}}{\rho_{00}}\hat{\boldsymbol{x}}\,, 
\\
&&{\nabla} \boldsymbol{\cdot} \boldsymbol{v} = 0\,,\qquad
\frac{\partial{\rho}}{\partial{t}} +  \boldsymbol{v} \boldsymbol{\cdot} {\nabla} \rho = D \nabla^2 \rho,
\end{eqnarray}
where $\rho_{00} = 1$~g/cm$^3$ is a reference density, $g$ is the gravitational acceleration, and $\nu = 0.01$~cm$^2$/s is the kinematic viscosity of fresh water.  The salt diffusivity $D = 2\times 10^{-5}$~cm$^2$/s is equal to the value for sodium chloride, which is used in the laboratory experiments described above, resulting in a Schmidt number of $\nu/D = 500$.  Given the large Schmidt number, the density field does not mix over the course of our simulations or experiments, which is expected given the lack of wave breaking and overturning for the parameters that we have examined.  The tidal flow $\boldsymbol{u}_{\mathrm{tide}} = -\hat{\boldsymbol{x}} A \omega \cos{\omega t}$ is driven by the tidal force $F_{\mathrm{tide}} = \rho_0 A \omega^2 \sin{\omega t}$, where a tidal excursion $A = 0.1$~cm matches the value used in the experiments.  The time step is chosen to correspond to 2000 time steps per period for the experiments that have an exponential stratification (described below), and 4000 time steps per period for the case with uniform stratification $(N = \mathrm{const})$  used to compare with analytical theory.\cite{smith03}   The simulations are run long enough to yield a steady-state for at least three periods, which typically requires at least 20 tidal periods.  

Two different stratifications are used in the numerical simulations, an exponential $N(z)$ to compare experiments and simulations and a constant $N= 1.55$~rad/s to compare simulations with analytical theory.  For the latter case, we choose a tidal frequency of $\omega = 0.255$~rad/s, which yields an internal wave beam slope of $S_{\mathrm{IW}} = \sqrt{\omega^2/(N^2-\omega^2)} = 1/6$.  

The computational grids are generated with Pointwise Gridgen.  Grid~I, tailored to match the the experiment (cf. Fig.\ \ref{fig:wsim_lab}), spans $-120 < x < 120$~cm and $0 < z < 55$~cm and has approximately $1.7 \times 10^6$ control volumes.  The  simulation topography is composed of a knife edge with height $H = 5$~cm and width $W/H = 0.032$ that is attached to a plate that extends nearly across the computational domain. The structured grid has smoothly varying spatial resolution with grid spacings of $\Delta x = 0.058$~cm and $\Delta z = 0.03$~cm near the topography, and $\Delta x = 2$~cm and $\Delta z = 0.1$~cm for locations far away from the topography.  To mimic the absorbing fiber mesh along the side boundaries in the experiments, we apply a Rayleigh damping term $(\propto \boldsymbol{v} - \boldsymbol{u}_{\mathrm{tide}})$ for $|x| > 50$~cm.  

The second domain, Grid~II, is designed to minimize finite-size effects to allow for comparisons with the analytical predictions of \citet{smith03}; for this case $N$ is constant as in the analytical theory.  Grid~II spans $-400 < x < 400$~cm and $0 < x < 80$~cm and has approximately $1.1 \times 10^6$ control volumes.  The knife edge in this case has the same dimensions as in the experiment, but the base of the knife edge is centered at $(x = 0, z = 0)$.  The grid spacing is $\Delta x = 0.02$~cm and $\Delta z = 0.02$~cm in the vicinity of the knife edge, and smoothly increases to $\Delta x = 2$~cm and $\Delta z = 0.15$~cm along the periphery.  Rayleigh damping is applied for $|x| > 300$~cm and $z > 50$~cm to prevent reflections.  

In the simulations for both Grid~I and for Grid~II,  no-slip boundaries conditions are applied along the topography, top boundary, and bottom boundary, while periodic boundary conditions are used in the $x$-direction.  Convergence tests with the spatial and temporal resolution doubled (halved) changed the computed velocities by less than 1\% (4\%).  

\begin{figure}
\includegraphics[width=\textwidth]{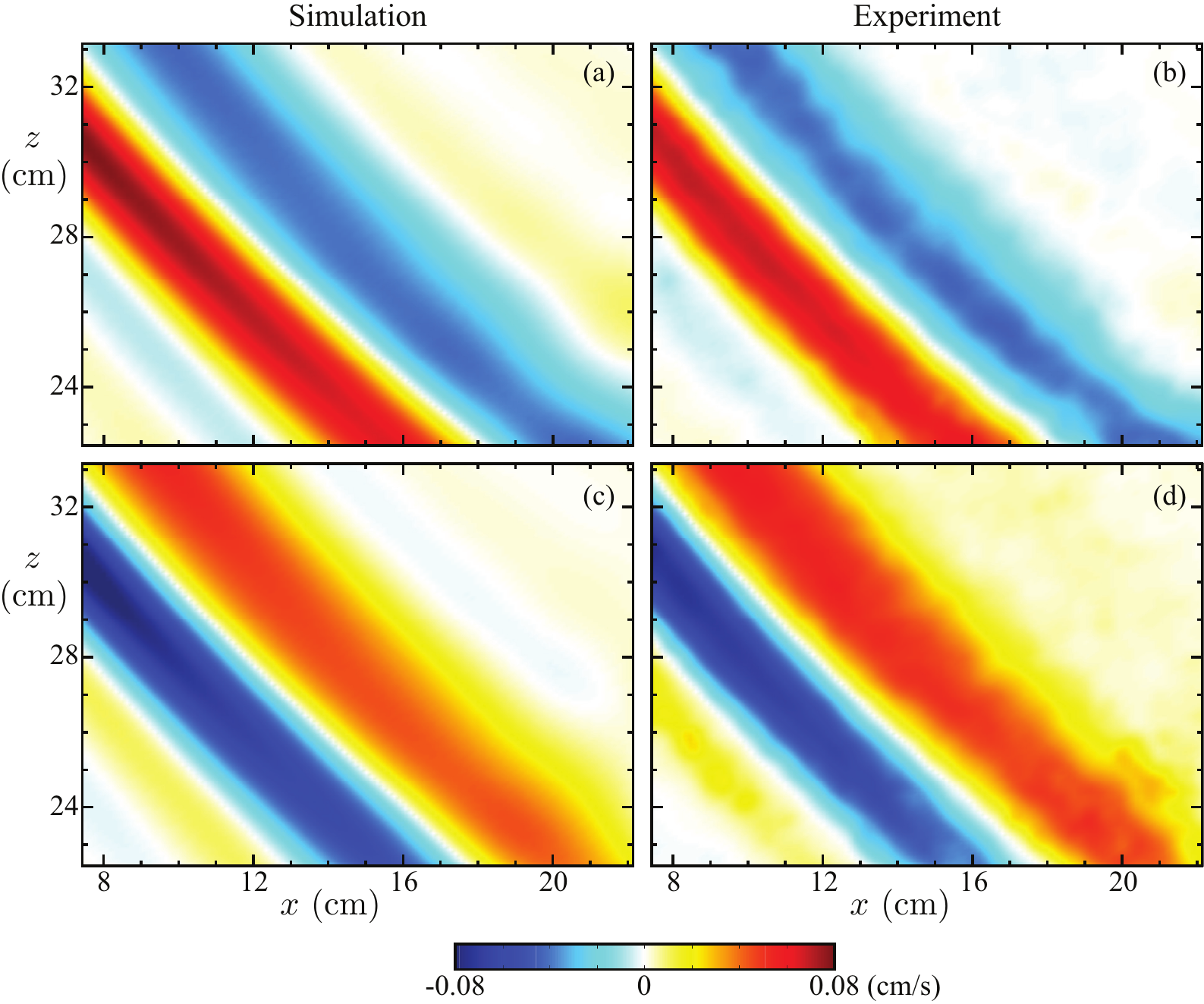}
\caption{Snapshots of the horizontal (top) and vertical (bottom) components of the velocity field (color) determined in simulation (left) and experiment (right) agree well.  The measurement region is shown as a dashed box in Fig.\ \ref{fig:wsim_lab}, and that figure's caption gives the conditions.}
\label{fig:uw_pcolor}
\end{figure}

\begin{figure}
\includegraphics[width=4.6181in]{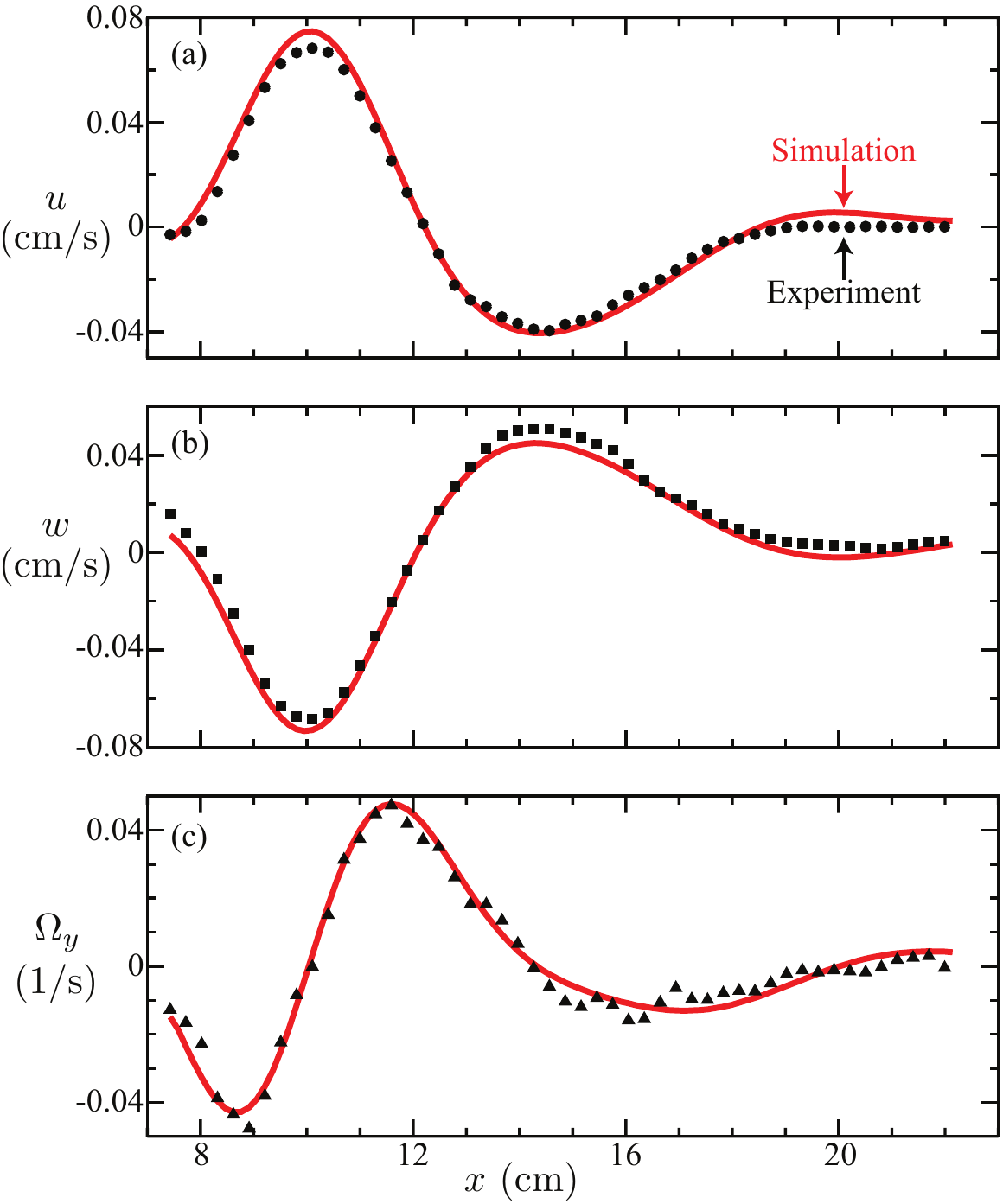}
\caption{Cross-sections of (a) the horizontal velocity component $u$, (b) vertical velocity $w$, and (c) vorticity $\Omega_y = ({\nabla} \times \boldsymbol{u})_y$ measured at $z = 27.7$~cm show excellent agreement between the experimental measurements (black symbols) and numerical simulations (solid (red) curves).  The conditions for the measurements and computations are given in the caption of Fig.\ \ref{fig:wsim_lab}.}
\label{fig:u_w_vort_vs_x}
\end{figure}

A snapshot of the vertical velocity field computed using Grid~I is shown in Fig.\ \ref{fig:wsim_lab}.  The knife edge (centered at $x = 0$) produces four internal waves beams, two that initially propagate upward before reflecting from the base plate, and two others that propagate downward. The edges of the base plate at $|x| = 39.4$~cm also produce weaker internal wave beams.  The area shown corresponds to the laboratory tank; the domain Grid~I for the simulations is much wider.  Rayleigh damping absorbs the wave beams outside of the area shown.  To validate the simulation results, we  compare in Fig.~\ref{fig:uw_pcolor} the computed velocity field with that measured in the laboratory experiments. The agreement is quite good, as found in our prior comparisons of results from experiments with simulations using the CDP code.\cite{king09,king10,paoletti12}  The quantitative agreement between simulation and experiment is illustrated by the cross-sections of the velocity and vorticity fields shown in Fig.\ \ref{fig:u_w_vort_vs_x}.  Similar agreement between simulation and experiment is found for other times and spatial locations.

\subsection{Experimental techniques}
\label{subsec:expt_techniques}

We  examine the generation and propagation of internal waves in a glass tank that spans $-45 < x < 45$~cm, $0 < y < 45$~cm, and $0 < z < 60$~cm. The topography is inverted with its base at $z = 45$~cm (see  Fig.\ \ref{fig:wsim_lab}). A knife edge ridge with a height $H = 5$~cm and  width (in the $x$-direction) $W/H = 0.032$ is centered at $x = 0$ and spans the tank in the $y$-direction.  The ridge is connected to a base that spans $-39.4 < x < 39.4$~cm, $1 < y < 44$~cm, and $45 < z  < 46.27$~cm, to give a no-slip boundary condition.  The edges of the base plate at $|x| = 39.4$~cm are rounded to reduce the spurious generation of internal tides from the ends.  

A buoyancy frequency varying exponentially with depth is chosen to model the deep ocean.\citep{king12,paoletti12} A density profile corresponding to exponentially varying buoyancy frequency is produced using the generalized double bucket method described by \citet{hill02}. The density as a function of depth is measured using an Anton Paar density meter; in the bottom of the tank the fluid density is 1098~kg/m$^3$, and at the top surface (55 cm above the bottom) the density is 1000~ kg/m$^3$.  The resultant buoyancy frequency profile is
\begin{equation}
N(z) = 1.87\exp(-0.0141z) \mathrm{~ rad/s}
\label{eq:expt_rho}
\end{equation}
over the range $0 < z < 50$~cm.  The buoyancy frequency at the base of the experimental topography is $N_{\mathrm{B}} = 0.99$~rad/s, and it exponentially increases towards its maximum value of 1.87~rad/s at $z = 0$~mm at the bottom of the tank.  

Tidal flow is  generated by oscillating the rigid topography and base plate rather than by driving the fluid over stationary topography.  Our velocity measurements, then, are in the reference frame of the tidal flow.  The position of the topography is given by
\begin{equation}
\label{eq:expt_tide}
x(t) = A[1-e^{(-2\omega t/3\pi)}]\sin{(\omega t)},
\end{equation}
where the tidal excursion is $A = 0.1$~cm and the tidal frequency is $\omega = 0.90$~rad/s.  The exponential term is added to allow for a gradual increase in the oscillation amplitude, which reaches 99\% of its peak value after approximately 3.5 tidal periods.\cite{echeverri09}  The Reynolds number based upon the topographic height and tidal flow is $Re = A\omega H/\nu = 48$, while the Froude number is $Fr = A\omega/N_{\mathrm{B}}H = 0.02$.  To minimize finite-size effects, we reduce reflections of the internal waves at the side boundaries by placing fiber mesh at $|x| = 45$~cm.  

We obtain two-dimensional velocity fields $\boldsymbol{v} = (u,w)$ by particle image velocimetry\cite{adrian91} in a vertical plane along the center of the tank at $y = 22.5$~cm.  Hollow glass spheres with diameters $8 < d < 12$~$\mu$m and densities in the range $1.05 < \rho < 1.15$~g/cm$^3$ serve as seed particles, and are illuminated by a 5~mm thick laser sheet with a wavelength of 532~nm and a power of 2~W.  We capture the motion of the tracer particles 40 times per period with a 12-bit CCD camera with $1296 \times 966$ pixel resolution spanning 15.25~cm in the $x$-direction and 11.36~cm in the $z$-direction, as shown schematically by the dashed box in Fig.\ \ref{fig:wsim_lab}.  We use the CIV algorithm developed by \citet{fincham00} to determine the instantaneous velocity fields, which are interpolated to a regular $100 \times 100$ grid with spatial resolution $\Delta x = 0.15$~cm and $\Delta z =0.11$~cm.

\section{Results} \label{sec:Results}

In Sec.~\ref{subsec:compare_prior_theory} we show, using velocity and pressure field data from a   direct numerical simulation, that our method for computing internal wave power from velocity data  alone yields results in good accord with the wave power computed in the usual way from velocity and pressure data. In the same section we compare the radiated power given by the analytical predictions of Llewellyn Smith and Young with the power computed in the direct numerical simulations. In Sec.~\ref{subsec:starting_pt} we examine how the radiated power computed from the velocity field depends on the starting point for the calculation of the stream function from the velocity data. In 
Sec.~\ref{subsec:comparing_expts_and_sims} laboratory measurements of a velocity field are used to compute energy flux, which is found to agree with results obtained from direct numerical simulations that give both velocity and pressure fields.   
\subsection{Internal wave power from fluxes $\langle{\bm{J}}_p\rangle$ 
and $\langle{\bm{J}}_{\psi}\rangle$}
\label{subsec:compare_prior_theory}

\begin{figure}[htbp]
\includegraphics[width=\textwidth]{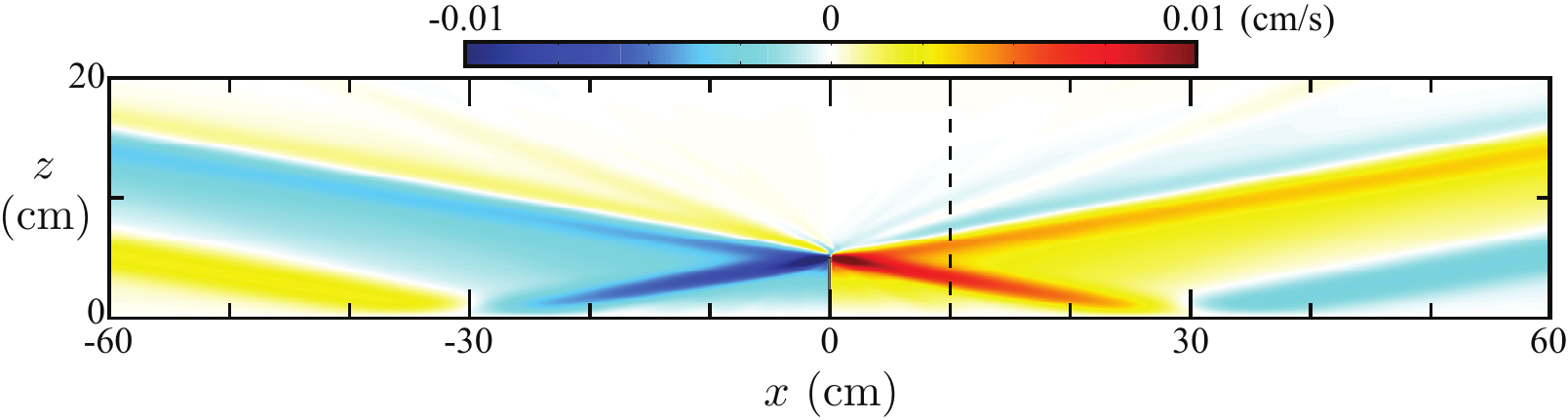}
\caption{A snapshot of the velocity field (color) from a numerical simulation of tidal flow over knife edge topography for a fluid with uniform stratification $(N = \mathrm{const})$.  Four internal wave beams are generated at the tip of the knife edge and propagate at constant angles. Measurements of the horizontal energy flux through a cross-section at $x = 10$~cm (dashed line) are shown in Figs.~\ref{fig:flux_p_z0=0}(a) and \ref{fig:flux_P_vary_z0}(a).}
\label{fig:wsim_UniformN}
\end{figure}

In this subsection we compare the power computed by the stream function method with the power computed from the velocity and pressure fields. We assume constant stratification ($N$=constant) in order to validate the stream function method by comparison with analytic theory.  We take $\rho_{00}$ to be the average value of the background density over the domain.  The geometry and a snapshot of the computed velocity field are shown in Fig.~\ref{fig:wsim_UniformN}. 

\begin{figure}
\includegraphics[width=4.6488in]{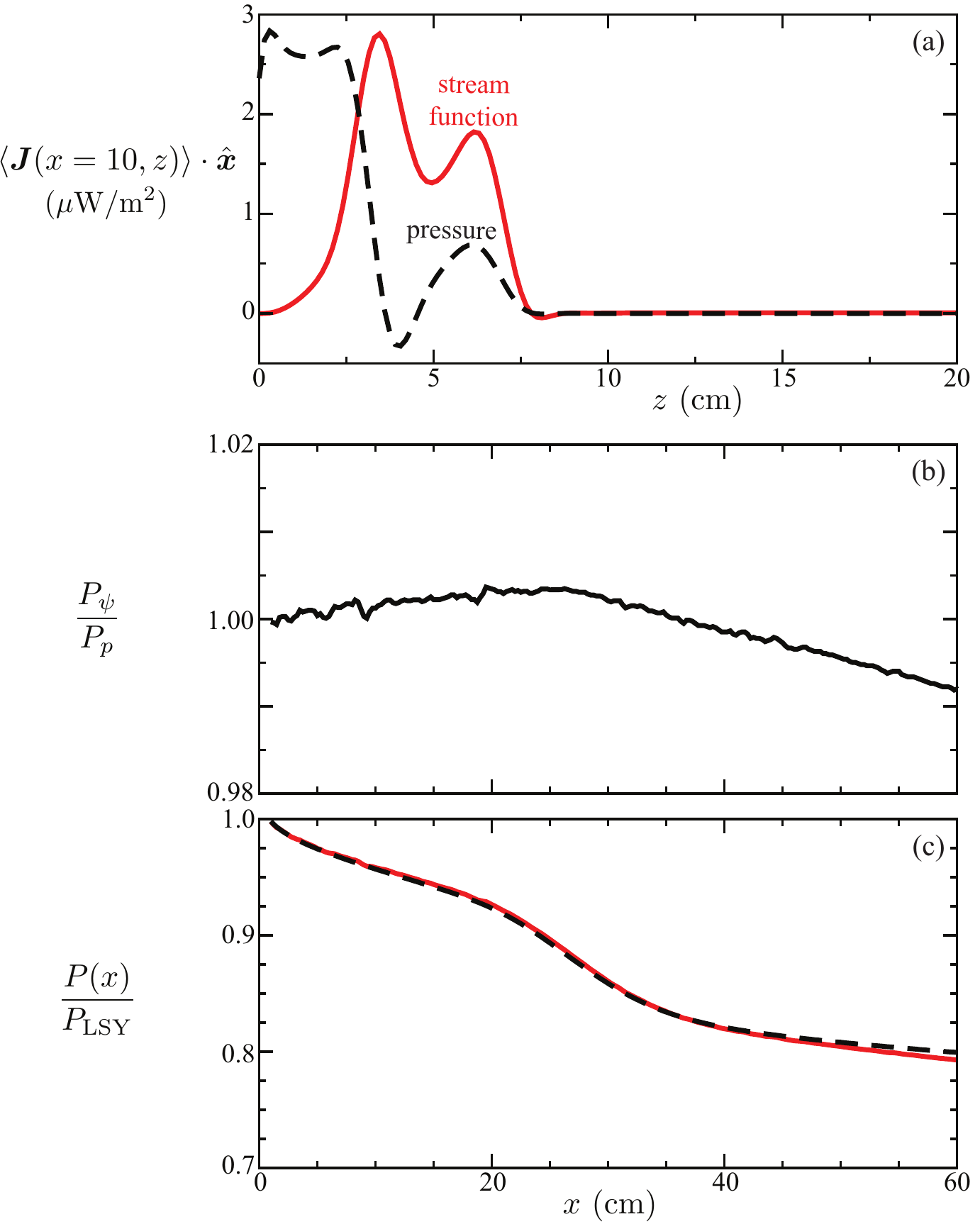}
\caption{(a) The horizontal energy flux $\langle \boldsymbol{J}_\psi \rangle \cdot \hat{\boldsymbol{x}}$ computed using the stream function method (Eq.~(\ref{j2}), red curve; $x_0 = 60$ cm, $z_0 = 0)$ differs from the energy flux 
$\langle \boldsymbol{J}_p \rangle \cdot \hat{\boldsymbol{x}}$ computed using the pressure method (Eq.~(\ref{j1}),   dashed line). The data are from the numerical simulation with $x$=10. (b) The total power $P_{\psi}$ computed by integrating $\langle \boldsymbol{J}_\psi \rangle \cdot \hat{\boldsymbol{x}}$ (solid (red) curve)  is in excellent agreement with the total power $P_p$ computed from the pressure and velocity, $\langle \boldsymbol{J}_p \rangle \cdot \hat{\boldsymbol{x}}$; the power is shown for vertical cross-sections at different $x$.  (c) The total radiated internal wave power obtained from numerical simulations  for both the stream function and velocity-pressure approaches are compared with the prediction of the  linear analysis of \citet{smith03}.  The computed power agrees with inviscid theory near the topography ($x=0$) but decreases with increasing $x$ due to viscous dissipation.}
\label{fig:flux_p_z0=0}
\end{figure}

The energy flux $ {\bm{J}}_p$ computed from the pressure in Eq.~(\ref{j1}) and the flux $ \boldsymbol{J}_\psi $ from the stream function in Eq.~(\ref{j2}) differ by  ${\nabla} \times \(\psi p \hat{\boldsymbol{y}}\)$, which represents a gauge transformation.  The striking difference between the  time averaged horizontal $\hat{\mathbf{x}}$ components of the two fluxes  is  illustrated  in Fig.~\ref{fig:flux_p_z0=0}(a). 

Even though the energy flux fields computed using the pressure and the stream function methods differ, as mentioned before, the radiated power should be the same because it is given by the volume integral of the divergence of the energy flux (cf.\  Eq.~\eqref{eq:J_integrals}).  Since the divergence of the gauge transformation term ${\nabla} \times \(\psi p \hat{\boldsymbol{y}}\)$ is zero, it does not contribute to the power.  Indeed, the radiated power computed from our simulation data by the stream function and pressure methods are in excellent agreement, as Fig.\ \ref{fig:flux_p_z0=0}(b) illustrates; the   root-mean-square difference between the two methods is less than 0.5\%. This is our main result: the radiated internal wave power can be determined using velocity field data alone. 
 
We now compare the computed radiated power with that predicted by \citet{smith03} for tidal flow of an inviscid, uniformly stratified fluid over knife edge topography in an infinitely deep ocean (in the absence of rotation),
\begin{equation}
P_{\mathrm{LSY}} = \frac{\pi}{4} \rho_0 H^2 A^2 \omega^2 \sqrt{N^2-\omega^2}L_y, 
\label{eq:P_LSY}
\end{equation}
where $L_y$ is the length of the topography in the direction orthogonal to both the tidal flow and gravitational acceleration.  We have replaced $N$ in Ref.~[\onlinecite{smith03}] with $\sqrt{N^2-\omega^2}$ to account for  nonhydrostatic effects.  The radiated internal wave power computed from  the stream function and pressure methods is compared to the inviscid theory prediction by using $P_{\mathrm{LSY}}$ as normalization in  Fig.~\ref{fig:flux_p_z0=0}(c). Immediately outside the laminar boundary layer at $x = 1$~cm, our computed values are 99.8\% of the value predicted by the inviscid theory.  Further away from the topography (increasing $x$), the power monotonically decreases owing to viscous dissipation, which is not present in theoretical studies.\cite{balmforth02,smith02,smith03,stlaurent03,khatiwala03,petrelis06,nycander06,lorenzo06,griffiths07,garrett07,balmforth09,echeverri10,zarroug10}  The power rapidly decreases near $x \approx 0$ from dissipation within the laminar boundary layer.  Near $x = 25$~cm the internal wave beams reflect from the bottom, producing a boundary layer with enhanced dissipation relative to the freely propagating internal waves in the bulk of the fluid (cf. Fig.~\ref{fig:wsim_UniformN}).   Although  viscosity was neglected in our derivation of the energy flux, the method seems to account for viscous dissipation quite well.

\subsection{Dependence on stream function starting point}
\label{subsec:starting_pt}

In order to compute the energy flux and radiated power  using only velocity data, the stream function must first be computed by using Eqs.~(\ref{eq:psi3}) and (\ref{eq:psi4}), which requires the choice of both a starting point $\(x_0,z_0\)$ and a value for the arbitrary integration constant $\psi\(x_0,z_0,t\)$.  \citet{balmforth02} effectively chose a point on the boundary and set $\psi\(x_0,z_0,t\) = \mathrm{constant}$, which is justified by the no-penetration boundary condition.  However,  experimental observations often do not include points on a solid boundary, and that is the case in our experiment  (see the  dashed box in Fig.~\ref{fig:wsim_lab}).   Therefore, as a substitute for  solid boundary points we choose effective boundary points  starting  as far away from the internal wave beams as possible, assuming that the stream function values at those points closely match those of the solid boundary and are thus constant in time.  Further, since the value of the constant itself does not change the flux, we choose $\psi\(x_0,z_0,t\) = 0$.  

To explore the effects of the choice of starting point on the calculation of the stream function, we consider internal waves generated by tidal flow of a uniformly stratified fluid $(N = \mathrm{constant})$ over a knife edge for the domain 2 (Grid II) described in Sec.~\ref{subsec:simulations}.  This domain, larger than the experimental domain (domain 1), removes the laboratory domain's finite-size effects and spurious generation of additional internal waves from the base plate.   The snapshot of the vertical velocity field in Fig.~\ref{fig:wsim_UniformN} shows the four internal wave beams that are generated by the knife edge with its base at $(x = 0, z = 0)$.  Two of the internal wave beams radiate upward, and two other beams initially propagate downward, reflect near $x = \pm 25$~cm, and then propagate upward.  The waves are absorbed by Rayleigh damping before reflecting from the boundaries.  

\begin{figure}
\includegraphics[width=5.3503in]{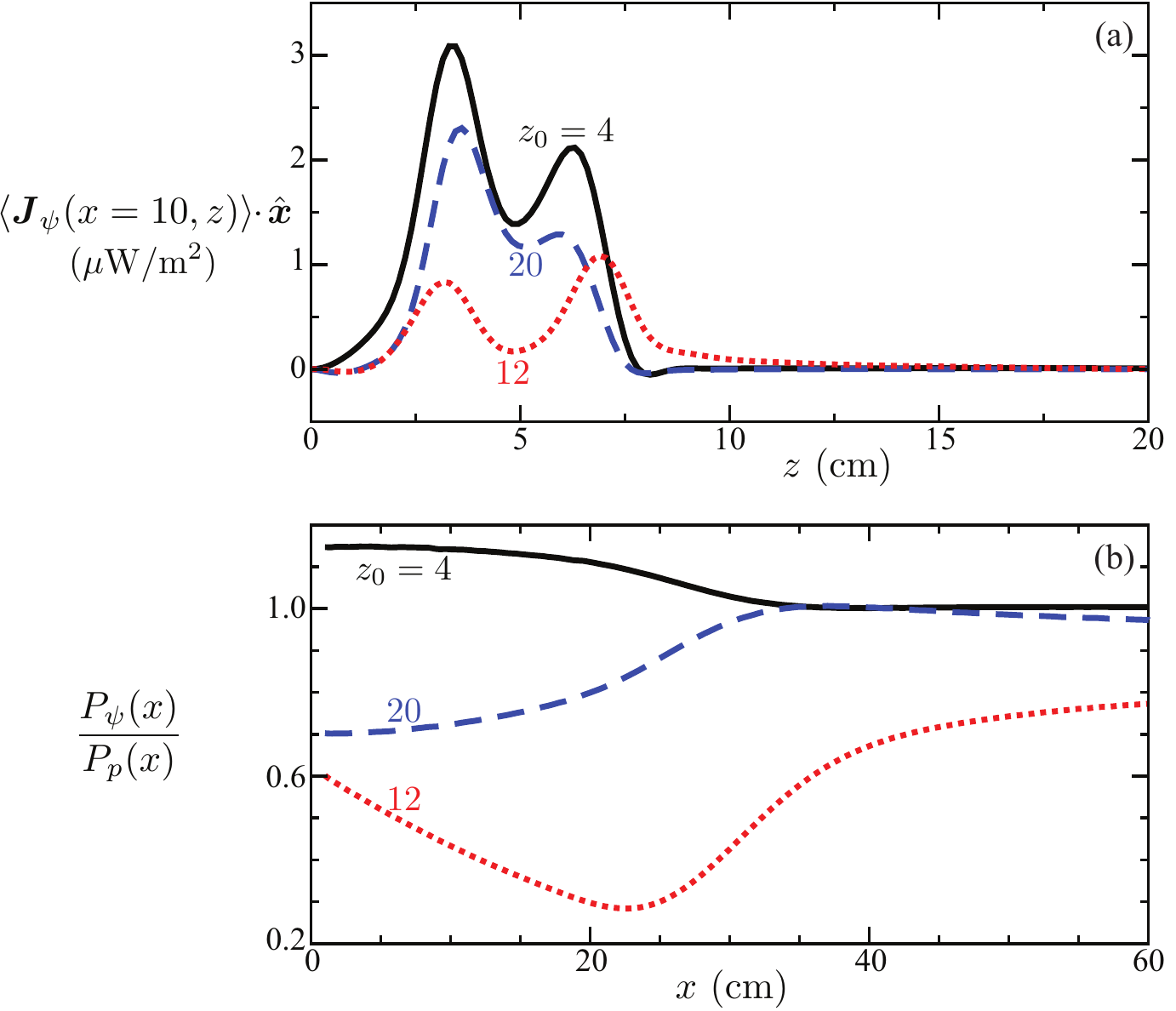}
\caption{(a) The horizontal energy flux $\left< \boldsymbol{J}_{\psi}\right> \cdot \hat{\boldsymbol{x}}$ determined by the stream function method (Eqs.\ (\ref{eq:psi3}) and (\ref{eq:psi4})) for starting points with $x_0 = 60$~cm and $z_0$ = 4, 12, and 20~cm. (b) The internal wave power $P_{\psi}$ obtained by integrating the horizontal flux vertically for different $x$, normalized by the power $P_p$ computed using the pressure method, for the three different stream function calculation starting points.}
\label{fig:flux_P_vary_z0} 
\end{figure}

The horizontal energy flux and the total radiated internal wave power are shown in Fig.~\ref{fig:flux_P_vary_z0} for three  starting points for the computation of the stream function (with $\psi\(x_0,z_0,t\) = 0$).  The horizontal energy fluxes computed from the  three representative starting points differ significantly; the starting point with $z_0 = 4$~cm is between the bottom boundary and the reflected wave that propagates to the right; the starting point with $z_0 = 12$~cm is between the two rightward-propagating internal waves; and the starting point with $z_0 = 20$~cm is above both internal waves but far from any solid boundary. The energy flux is strongest for $z_0 = 4$~cm. The energy flux has a similar structure for $z_0 = 20$~cm, but the  flux is much lower for $z_0 = 12$~cm.

The total radiated power $P_{\psi}$ integrated for vertical cross-sections at different $x$ is shown in 
Fig.~\ref{fig:flux_P_vary_z0}(b) for the three different starting points of the stream function calculation. $P_{\psi}$  is normalized by the power computed by the pressure method, $P_p$. For $x > 30$~cm, the power $P_{\psi}$ computed for starting points outside of the internal wave beams ($z_0 = 4$ and $z= 20$~cm) is in excellent agreement with $P_p$; the rms difference is  0.5\% for $z_0 = 4$~cm and 1\% for $z_0 = 20$~cm.  For $x < 30$~cm (i.e., farther away from the $x_0=60$ cm starting point), $P_{\psi}$ computed with $z_0 = 4$~cm is larger than $P_p$ by as much as 15\%, and for $z_0 = 20$~cm, $P_{\psi}$ is smaller than $P_p$ by as much as 30\%.  For the starting point located between the internal wave beams ($z_0 = 12$~cm) (cf.\  Fig.~\ref{fig:wsim_UniformN}), $P_{\psi}$ is smaller than $P_p$ by at least 20\% and as much as 70\%.   This example illustrates that the starting point for a stream function calculation of the flux should be 
outside of the internal wave beams, and the total internal wave power should be obtained for cross-sections far enough away from the topography to avoid near-field effects and close enough to the starting point for the stream function to reduce the cumulative error from quadrature over long paths.

\subsection{Comparison of experiment and numerical simulation}
\label{subsec:comparing_expts_and_sims}

\begin{figure}
\includegraphics[width=5.5946in]{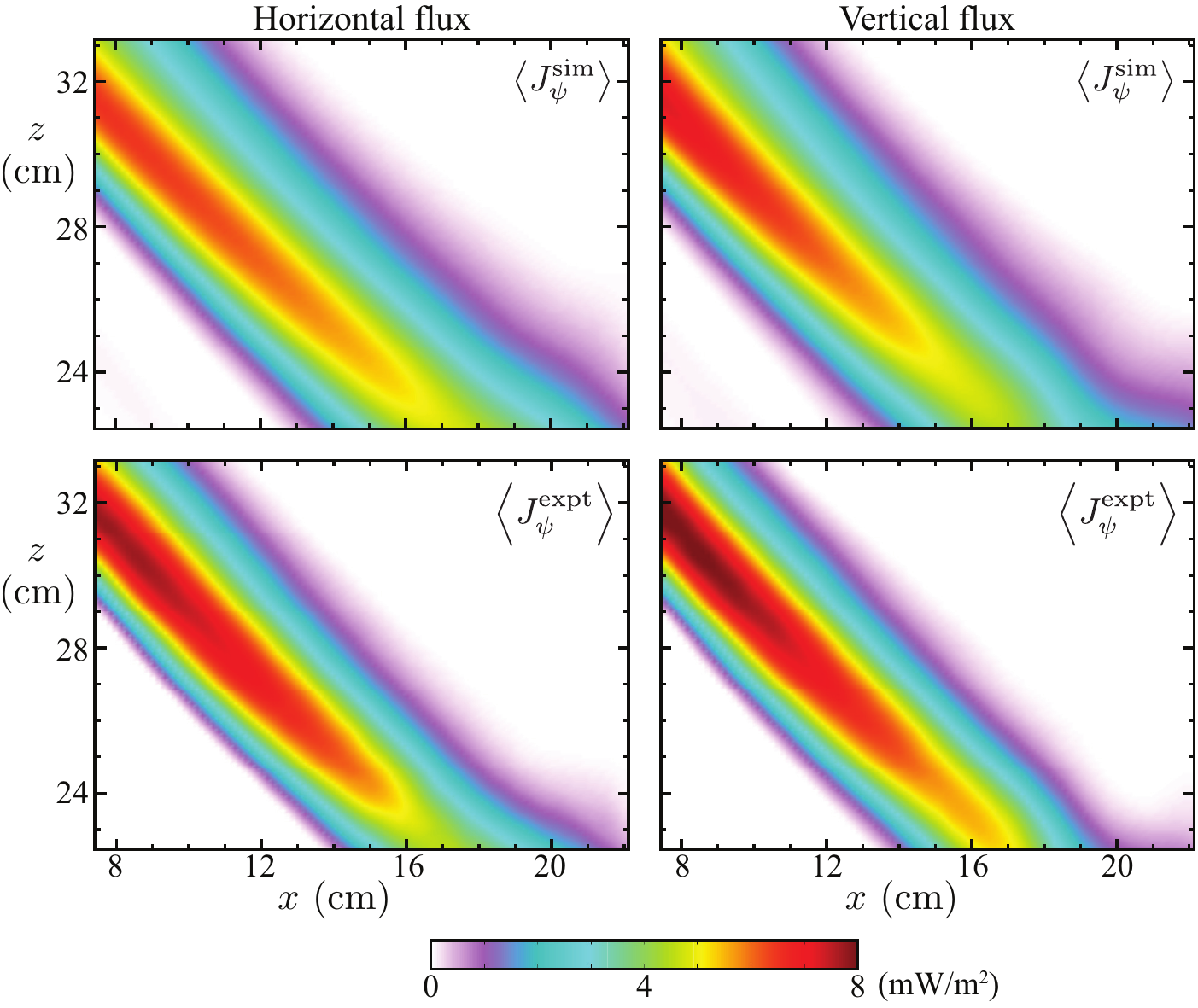}
\caption{The tidally averaged horizontal (left) and vertical (right)  energy flux computed  by the stream function method for  simulation data (top),   compared with the the method applied to laboratory data (bottom).  The region shown  is indicated by the  dashed rectangle in Fig.\ \ref{fig:wsim_lab}.}
\label{fig:flux_pcolor}
\end{figure}

Figure \ref{fig:flux_pcolor} compares the energy flux field from the numerical simulations $\langle\boldsymbol{J}_{\psi}^{\mathrm{sim}}\rangle$ with that  from a laboratory experiment $\langle\boldsymbol{J}_{\psi}^{\mathrm{expt}}\rangle$ for the same conditions.   
In this figure it is seen that the energy flux computed by the stream function method for  the simulation and laboratory agree well.


The  radiated internal wave power computed for the simulation data by integrating the energy flux across the beam is 3.09~nW (per cm of topography) and 3.01~nW, respectively for the integrals of $\langle\boldsymbol{J}_p^{\mathrm{sim}}\rangle$ and $\langle\boldsymbol{J}_{\psi}^{\mathrm{sim}}\rangle$ across the beam; the internal wave power obtained by integrating the energy flux obtained from the laboratory data is 2.83~nW. The difference between the experimental and simulation results for the radiated power arises from differences between the laboratory system and the simulation rather than from errors in the stream function methodology, which has been validated by using pressure and velocity data from the numerical simulation.  The differences between the experiment and simulation include the viscosity, which is constant in the simulation but varies in the experiment by 20\% from the tank bottom to the fluid surface; the sidewall boundary condition, which in the laboratory tank is absorbing because the walls are  lined with a fiber mesh to reduce reflections; and the shape of the ends of the base plate on which the topography was mounted. Despite these differences the agreement is within 10\%. 

\section{Discussion} \label{sec:Discussion}
The method presented for determining energy flux and radiated power for internal waves using only velocity field data could provide  opportunities for laboratory experiments and field measurements that go beyond the capabilities of existing theory. While theoretical \cite{kistovich98} and experimental \cite{paoletti12}  studies have examined the viscous decay of the velocity field for propagating internal waves in arbitrary stratifications, theoretical studies of internal wave  generation for flow over topography have been for  inviscid fluids.\cite{bell75,balmforth02,smith03,khatiwala03,petrelis06,nycander06,balmforth09,echeverri10}
Figure \ref{fig:flux_p_z0=0}(c)  shows that the the stream function method yields the decay of the wave power as well as the generated power. Therefore, velocity measurements can be used to characterize both the conversion of tidal motions to internal waves and the viscous decay as the waves propagate away from the topography.

Theoretical studies of the conversion of tidal motions to internal wave power have focused on laminar flow over the topography, but the boundary currents can become intense and unstable, particularly for critical topography where the slope of the topography is equal to the local slope of the internal wave beams. \cite{zhang08,gayen10,gayen11a,gayen11b,dettner13} Indeed, recent numerical simulations have found that turbulence generated near critical topography can reduce the radiated internal wave power.\cite{rapaka13} While the turbulence is 3-dimensional, the far field internal beams can be predominantly 2-dimensional\citep{aguilar06} and hence could be determined by the stream function method.  

The energy flux field computed from the stream function method also offers a different perspective on the underlying physics: the  flux computed from the pressure and stream function methods differ, as Fig.\ \ref{fig:flux_p_z0=0}(a) illustrates. The energy flux computed by the pressure method can be interpreted as \textit{input} power since it is strongest for values of $z$ less than the height of the topography ($H = 5$~cm), where viscous drag is large. On the other hand, the energy flux computed by the stream function method is peaked at the center of the internal wave beams 
(cf.~Fig.~\ref{fig:wsim_UniformN}).  Thus, one might  interpret the horizontal energy flux measured by the stream function method as corresponding to the power \textit{output} by the radiated internal waves.  The input and output powers are found to be equal, as expected (Fig.~\ref{fig:flux_p_z0=0}(b)). 
\section{Conclusions} \label{sec:Conclusions}
We have shown that the energy flux and the integrated wave power power for 2-dimensional internal waves can be determined using  knowledge of only the velocity field, which can be written in terms of a single scalar field,  the stream function. The energy flux field and radiated power can be computed from Eqs.\ \eqref{j2tavg} and \eqref{eq:J_integrals}, in analogy with the methods used in prior theoretical work.\cite{balmforth02,smith03,khatiwala03,petrelis06,nycander06,balmforth09,echeverri10}  We have tested the stream function method for determining internal wave flux and power using results obtained for tidal flow over a knife edge, computed with a numerical simulation code that has been validated in previous studies.\cite{king09,king10,paoletti12,dettner13} The results for the radiated internal wave power obtained from the stream function and pressure methods are found to agree within one percent,  {\em if} the starting point for the stream function calculation is chosen near a boundary or far from the internal wave beams.  We also made laboratory measurements of the velocity field for tidal flow past a knife edge and used those data to determine the internal wave power, which agreed with the numerical simulation results within ten percent. Given the excellent agreement between the results from the pressure and stream function approaches for the simulation data, we believe the agreement between the experiment and simulation could be improved by designing an experiment that better satisfied the assumptions of the simulations.

We have submitted  Supplementary Material that provides a Matlab code with a graphical user interface for the stream function method of determining energy flux and internal wave power from 2-dimensional velocity field data. A step-by-step description of the algorithm and its implementation are also included in the Appendix \ref{appendix}

\begin{acknowledgments}
We thank Bruce Rodenborn for help with the code and the GUI, and Likun Zhang and Robert Moser for helpful discussions. The computations were done at the Texas Advanced Computing Center. MSP and HLS were supported by the Office of Naval Research MURI Grant N000141110701, while  PJM and FML were supported by U.S.~Dept.\ of Energy Contract \# DE-FG05-80ET-53088. 
\end{acknowledgments}

\appendix

\section{Guide to Supplementary Material -- A  GUI for $\langle J_{\psi}\rangle$}
\label{appendix}

 A Matlab code and GUI for the stream function method for  determining the energy flux and power are available at the following URLs: \url{http://chaos.utexas.edu/wp-uploads/2013/12/internalwaves_streamfunction_fluxfield.zip} and \url{http://www.mathworks.com/matlabcentral/fileexchange/44833}. This appendix contains  information that is needed to use the GUI.

\subsection*{Input Data Format}

The user must first supply the .mat file which contains the velocity components, the grid, and a fluid parameters array containing the background density and buoyancy frequency information.
The names of the various arrays can be user-specified, but the defaults are as follows. Horizontal velocity: $u$, vertical velocity: $w$, horizontal coordinate: $x$, vertical coordinate: $z$, fluid parameters: $h\_rho0\_N$.  

\medskip

\underline{Velocity components}: The velocity components must be two separate arrays of identical shape. The first dimension is the $z$ direction, the second dimension is the $x$ direction, and the third is time. The units for the inputs for the program are cgs. 

\smallskip

\underline{Coordinate arrays}: The coordinate arrays must be in the same shape as the velocity components minus the time dimension and must also be separate arrays for the $x$ and $z$ coordinates. The arrays are in the form of outputs for the Matlab function ``meshgrid.'' Refer to the Matlab help documents for further details. 

\smallskip

\underline{Fluid parameters}: The fluid parameter array should contain as its first column the heights at which the background density (second column) and buoyancy frequencies (third column) are evaluated. The heights need not match with the $z$-component coordinate array specified previously; the values for the background density and buoyancy frequency will be interpolated (cubic) to fit it. If the Boussinesq approximation with uniform reference density and $N$ is being used, the two values can be input as scalars.

\subsection*{Other Parameters}

\begin{figure}
\includegraphics[width=\textwidth]{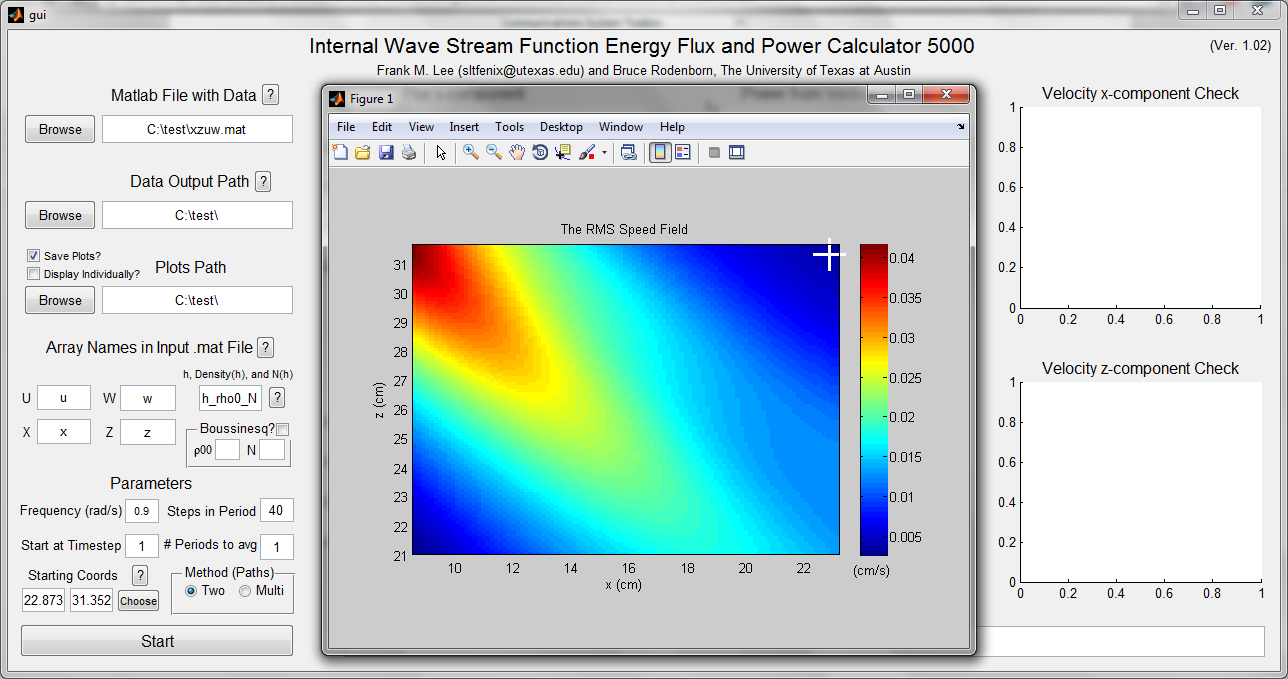}
\caption{The user can specify the starting point of the stream function calculation by clicking on a plot of the RMS speed field of the input data.}
\label{fig:gui1}
\end{figure}

The user then specifies the relevant parameters. The frequency of the internal wave field must be supplied in rad/s. Additionally, the number of timesteps in a period of oscillation, the timestep at which to start evaluating the power, and how many periods to average over must be specified. The energy flux expression is time-averaged over an integer number of periods. Additionally, the starting coordinates (in cm) must be specified for the stream function calculation, which can be chosen by clicking on a displayed plot of the velocity amplitude field, as shown in Fig.\ \ref{fig:gui1}. The stream function is taken to be zero at those coordinates at all times. The user can also choose between the two-path and multi-path methods. The multi-path method is roughly an order of magnitude slower than the two-path method, and should be used to reduce the error if the data supplied has a lot of noise. 

\subsection*{Calculation of the Stream Function}

Once all the data and parameters are supplied, the algorithm uses trapezoidal quadrature of the $x$-velocity values along the $z$-coordinates, and the $z$-velocity values along the $x$-coordinates to find the stream function at each grid point. For the two-path method, it will average over two simple $L$-shaped paths from the starting point to the evaluation point given by Eqs.~(16)
~and (17)
. For the multi-path method it will average over every $Z$-shaped path within the box that forms between the starting point and the evaluation point which is given by Eqs.~(18)
~and (19)
. Note that if the starting point and evaluation point have the same $x$ or $z$ coordinate, then the only possible path is a straight line. The two-path method calculates only two path integrals for each grid point (excluding the points in line with the starting point), which means it will integrate over $2  M  N - M - N$ paths, where $M$ is the grid size in $x$, and $N$ is the grid size in $z$. The multi-path method calculates $M + N + 2$ paths for each point, where $M$ and $N$ are the number of grid points between the evaluation point and the initial point in the $x$ and $z$ directions. Then the total number of paths integrated for the whole grid is $\frac{1}{2}[ M^2 N + M N^2 - ( M+N )^2 + 3 ( M+N ) - 4 ]$. The stream function is found for every timestep in the specified range. Derivatives of the calculated stream function are taken and checked against the input velocity components at the initial timestep at the middle of the domain.

\subsection*{Calculation of the Energy Flux}

\begin{figure}
\includegraphics[width=\textwidth]{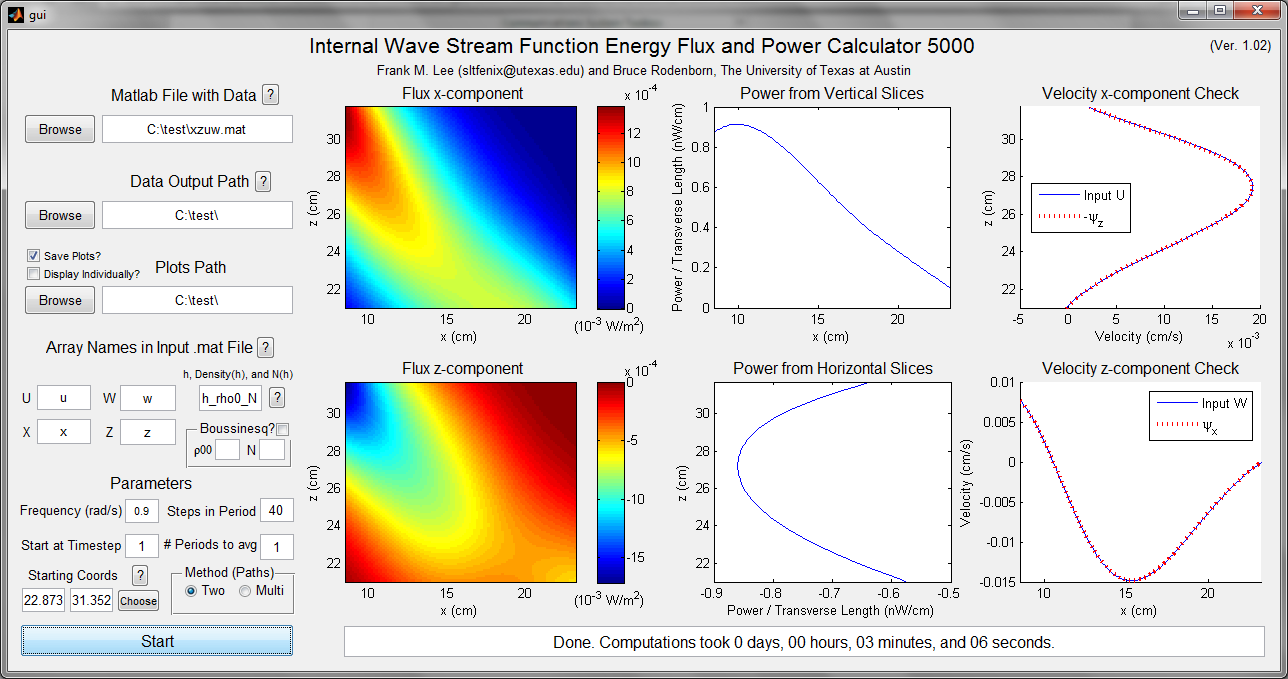}
\caption{After the various input parameters are inserted  into the GUI, the Matlab program calculates and displays the flux fields, the powers, and velocity checks.}
\label{fig:gui2}
\end{figure}

Once the stream function $\psi(x,z,t)$ has been calculated, $\varphi(x,z)$ and its derivatives are calculated (Eqs.~(20)
~--  (22)
). The real part of $\varphi$ is found by trapezoidal quadrature in the time direction at each grid point where the integrand is the product of the stream function and $\cos \omega t$. The imaginary part is found using $\sin \omega t$ in place of $\cos \omega t$. The derivatives are done the same way except the velocity components are used instead of the stream function. Then the energy flux (Eq.~(13)
) is calculated using these quantities. The flux fields and the powers are displayed (Fig.\ \ref{fig:gui2}) and output into both .txt and .mat files to the specified folder.

\end{document}